# An Analytical Solution for the Wigner-Boltzmann Transport Equation in the Relaxation Time Approach

A. R. Fernandes Nt.

*Centro Federal de Educação Tecnológica –CEFET/RJ, CEP 23.953-030, Angra dos Reis, Brazil*

The quantum version of the Boltzmann transport equation (Wigner-Boltzmann equation) is a quite useful tool to investigate the effects of energy dissipation in quantum systems. Numerical approaches uses to be employed in order to stablish a suitable solution. In this paper, an analytical solution is shown to exist when the constant relaxation time approximation is considered. The formalism presented here is capable to avoid some unphysical features early reported in literature for the conventional boundary condition scheme.

**INTRODUCTION**

The Boltzmann equation consists in a powerful way to describe systems out of thermodynamic equilibrium [1,2]. Generally, it involves the diffusion or transport of mass, energy or charge that makes the system evolve towards its equilibrium configuration, providing a suitable formalism to calculate a number of properties of a given physical system, such as thermal and electrical conductivity, diffusion coefficient, etc.

The equation is usually solved for a function, $f(\vec{r}, \vec{p})$, which gives the number of particles of a given system contained in a certain differential volume $(\vec{r} + d\vec{r}, \vec{p} + d\vec{p})$ of the phase space, consisting, therefore, in a distribution function. Such a definition, however, would not be a formal approach if quantum systems are considered, due to the uncertainty principle and the possibility of superposition of quantum states.

Meanwhile, a number of formulations of quantum mechanics in the phase space has been presented in the literature [3], allowing the definition of a quantum distribution function. The phase space formulation of quantum mechanics provides a straightforward way to establish a link between quantum and classical worlds, associating a quantum state to a function of position and momentum. The most widely known formulation of phase space quantum mechanics is the Weyl-Wigner representation [3-6]. This formulation is quite useful to describe quantum transport process which is important for condensed matter physics [4] and to the understanding of transition for classical statistical physics [5].

This work is divided in two parts. In the first one, the Weyl-Wigner formulation of quantum mechanics in the phase space is briefly presented and used to stablish the Wigner-Boltzmann equation, i.e., an equation describing the evolution of a phase space "quasi-distribution" function in the presence of collisions [7,8].

In the second part, an analytical solution is shown to exist when the relaxation time approximation is considered. For sake of illustration, the explicit solution for a beam of free particle crossing a dissipative region is written. Such a solution presents a delta function with complex argument in the entanglement term, which is call the generalized delta function [9], whose effect is to replace a real variable for a complex one. The potential of the presented formalism to avoid some unphysical features early reported in literature for the conventional boundary condition scheme [10] is also discussed.

## I. QUANTUM MECHANICS IN THE PHASE SPACE

**The Weyl-Wigner Representation**

Let $\psi(\vec{r})$ be the wave function associated to the quantum state $|\psi\rangle$ in the configuration space. The Wigner function $w_\psi(\vec{r},\vec{p})$ associated to $|\psi\rangle$ is defined as

$$w_\psi(\vec{r},\vec{p}) = \frac{1}{(2\pi\hbar)^3} \int d\vec{s}\, \psi^*\left(\vec{r} - \frac{\vec{s}}{2}\right) \psi\left(\vec{r} + \frac{\vec{s}}{2}\right) e^{-i\frac{\vec{s}\cdot\vec{p}}{\hbar}}, \tag{1}$$

which presents the following properties:

$$\int d\vec{p}\, w_\psi(\vec{r},\vec{p}) = |\psi(\vec{r})|^2, \tag{2a}$$

$$\int d\vec{r}\, w_\psi(\vec{r},\vec{p}) = |\tilde{\psi}(\vec{p})|^2, \tag{2b}$$

$$\int d\vec{r}\, d\vec{p}\, w_\psi(\vec{r},\vec{p}) = 1, \tag{2c}$$

where $|\psi(\vec{r})|^2$ and $|\tilde{\psi}(\vec{p})|^2$ are the probability distributions in the configuration and momentum space, respectively (i.e. $\tilde{\psi}(\vec{p})$ is the momentum wave function, obtained by the Fourier transform of $\psi(\vec{r})$). $w_\psi(\vec{r},\vec{p})$ is called a quasi-probability distribution function, since its physical meaning remains only in its integral over the configuration or momentum variables – so that it is a marginal distribution - and the integral in Eq. (1) can give negative values, so that Wigner functions cannot be viewed as a conventional probability distribution function.

Once a density matrix $\boldsymbol{\rho}_\psi = |\psi\rangle\langle\psi|$ is determined for a quantum state, the Wigner function can be defined as

$$w_\psi(\vec{r},\vec{p}) = Tr\left[\boldsymbol{\rho}_\psi \mathbf{P}_{ps}(\vec{r},\vec{p})\right]. \tag{3}$$

where

$$\mathbf{P}_{ps}(\vec{r},\vec{p}) = \frac{1}{(2\pi\hbar)^6} \int d\vec{s}d\vec{q}\; e^{\frac{i}{\hbar}\left[\vec{q}\cdot(\mathbf{\vec{r}}-\vec{r}) + \vec{s}\cdot(\mathbf{\vec{p}}-\vec{p})\right]}. \tag{4}$$

Is the Weyl-Wigner operator, which is an operator in the Hilbert space containing all key characteristics necessary to perform the phase space formulation of quantum mechanics. It is important since Hilbert space provides an easier and practical way to formulate quantum problems while, on the other hand, we shall be focused on describing certain physical phenomena (as diffusive electronic transport) in the phase space [11].

In fact, any operator $\mathbf{A}$ of the Hilbert space can be represented in the phase space by

$$A_{ps}(\vec{r},\vec{p}) = (2\pi\hbar)^3 Tr\left[\mathbf{A}\mathbf{P}_{ps}(\vec{r},\vec{p})\right]. \tag{5}$$

Inversely, it is easy to show that

$$\mathbf{A} = \int d\vec{r} d\vec{p}\ A_{ps}(\vec{r},\vec{p}) \mathbf{P}_{ps}(\vec{r},\vec{p}), \tag{6}$$

Generally, an operator **A** can be understood as an operatorial function $A(\vec{\mathbf{r}},\vec{\mathbf{p}})$. Considering the non-commutativity of position and momentum operators, the phase space function $A_{ps}(\vec{r},\vec{p})$ should not be obtained by a simple replacement $(\vec{\mathbf{r}},\vec{\mathbf{p}}) \to (\vec{r},\vec{p})$ since operator variables in different orders correspond to distinct operatorial functions. Transformation Eq. (5) maps each $A(\vec{\mathbf{r}},\vec{\mathbf{p}})$ function in the Hilbert space to a single correspondent $A_{ps}(\vec{r},\vec{p})$ function in the phase space.

Inversely, transformation Eq. (6) maps each $A_{ps}(\vec{r},\vec{p})$ function in the phase space to a single correspondent $A(\vec{\mathbf{r}},\vec{\mathbf{p}})$ function in the Hilbert space. This expression represents the Weyl correspondence [12].

The next step is to examine the case in which one have a product between two operators **A** and **B**, or $\mathbf{AB} = A(\vec{\mathbf{r}},\vec{\mathbf{p}}) B(\vec{\mathbf{r}},\vec{\mathbf{p}})$. In the phase space representation, by using Eqs. (4) and (5), we have

$$[AB]_{ps}(\vec{r},\vec{p}) = \frac{1}{(\pi\hbar)^6} \int d\vec{r}'d\vec{r}''d\vec{p}'d\vec{p}'' A_{ps}(\vec{r}',\vec{p}') B_{ps}(\vec{r}'',\vec{p}'') e^{\frac{2i}{\hbar}[(\vec{r}-\vec{r}')(\vec{p}-\vec{p}'')-(\vec{r}-\vec{r}'')(\vec{p}-\vec{p}')]}. \tag{7}$$

Note that $[AB]_{ps}(\vec{r},\vec{p})$ does not necessarily equals $[BA]_{ps}(\vec{r},\vec{p})$, which shows that transformation Eq. (7) ensures the non-commutative feature of quantum mechanics. The commutation operation in the phase space yields

$$[[A,B]]_{ps}(\vec{r},\vec{p}) =$$

$$\frac{1}{(\pi\hbar)^6} \int d\vec{r}'d\vec{r}''d\vec{p}'d\vec{p}'' A_{ps}(\vec{r}',\vec{p}') B_{ps}(\vec{r}'',\vec{p}'') \left( e^{\frac{2i}{\hbar}[(\vec{r}-\vec{r}')(\vec{p}-\vec{p}'')-(\vec{r}-\vec{r}'')(\vec{p}-\vec{p}')]} - e^{-\frac{2i}{\hbar}[(\vec{r}-\vec{r}')(\vec{p}-\vec{p}'')-(\vec{r}-\vec{r}'')(\vec{p}-\vec{p}')]} \right) \tag{8}$$

It is easy to show that first-order terms in $i\hbar$ on the right side of Eq. (8) correspond to the Poisson brackets. This fact shows a way by which classical limits for quantum mechanics can be obtained.

**The Wigner-Boltzmann Equation**

The time evolution for the density matrix $\boldsymbol{\rho}$ of a given quantum system described by the Hamiltonian operator $\mathbf{H} = (1/2m)\mathbf{p}^2 + V(\mathbf{z})$, where $V(\mathbf{z})$ is a position dependent potential, is given by

$$\frac{d\boldsymbol{\rho}}{dt} = \frac{1}{i\hbar}[\mathbf{H}, \boldsymbol{\rho}] + \frac{\partial \boldsymbol{\rho}}{\partial t}. \tag{9}$$

By taking the trace with the one-dimensional version of the Weyl-Wigner operator $\mathbf{P}_{ps}(z, p)$, given by Eq. (4), in both sides of Eq. (9), we have, in view of Eq. (8)

$$\frac{d}{dt} f(z, p) = \frac{p}{m} \frac{\partial}{\partial z} f(z, p) - \int dp' V_W(z, p - p') f(z, p') + \frac{\partial}{\partial t} f(z, p), \tag{10}$$

where $V_W(z, p - p')$ is the Wigner potential defined as

$$V_W(z, p - p') = \frac{2}{\pi \hbar^2} \int dz' V_{ps}(z') \sin\left(\frac{2}{\hbar}(z - z')(p - p')\right), \tag{11}$$

with $V_{ps}(z)$ being related to $V(\mathbf{z})$ as defined in Eq. (5). It is interesting to note that, in this case, the functional form of $V(\mathbf{z})$ is preserved, since it is a function of position operator only. In this case, we can simply write $V_{ps}(z) = V(z)$.

The connexion with Boltzmann formalism comes when, in addition to the effect of the potential, the effect of collisions is considered as well, so that $df(z, p)/dt = (\partial f(z, p)/\partial t)_{coll}$, being the term on the right the Boltzmann collision integral, so that Eq. (10) gives us

$$\frac{p}{m}\frac{\partial}{\partial z}f(z,p)-\int dp' V_W(z,p-p')f(z,p')+\frac{\partial}{\partial t}f(z,p)=\left(\frac{\partial}{\partial t}f(z,p)\right)_{coll}. \qquad (12)$$

Eq. (12) is known as the Wigner-Boltzmann equation.

The function $f(z,p)$ is given by a statistical sum of many generalized Wigner functions as defined in Eq. (3), so that only its integrals have an effective physical significance, i.e. its integral over momentum space gives the average number of particles at the interval between $z$ and $z+dz$, meanwhile its integral over position space gives the average number of particles with momentum at the interval between $p$ and $p+dp$. The integration over the entire phase space gives us the total number of particles in the system.

## II. DIFFUSIVE ELECTRONIC TRANSPORT

Let us dwell with a stationary situation ($\partial f(z,p)/\partial t=0$) in which we can write an equilibrium quasi distribution $f_{eq}(z,p)=\Sigma_k w_k(z,p)n_{eq}(\varepsilon_k)$, so that, according to Boltzmann approach, the effect of collision vanishes. In this case, Eq. (12) yields

$$\frac{p}{m}\frac{\partial}{\partial z}f_{eq}(z,p)-\int dp' V_W(z,p-p')f(z,p')=0. \qquad (13)$$

The one particle solution of Eq. (13) can be obtained by knowing the wave function satisfying the Schrodinger equation for the potential $V(\mathbf{z})$, and inserting it in Eq. (1). For a general quantum particle system, the solution can be obtained from the corresponding density matrix by inserting it in Eq. (3). Therefore, that is the simplest way given the difficulties in deal with the integral term in Eq. (13). In this case, we also establish $n_{eq}(\varepsilon_k)$ as a thermal energy state distribution.

In order to find a general solution for Eq. (12), let us set $f(z,p) = f_{eq}(z,p) + g(z,p)$, being $g(z,p) = \Sigma_k w_k(z,p) n(\varepsilon_k)$ the part of $f(z,p)$ which effectively represents the nonequilibrium contribution, so that we can use the relaxation time approximation as follows

$$\left(\frac{\partial}{\partial t} f(z,p)\right)_{coll} = \left(\frac{\partial}{\partial t} g(z,p)\right)_{coll} = -\sum_k \frac{1}{\tau_k} w_k(z,p) n(\varepsilon_k), \quad (14)$$

where $\tau_k$ is the meantime between collisions (relaxation time) for a particle in the state $k$. A usual approach consists in considering the relaxation time as a constant, representing the mean time between collisions for an electron at the Fermi level. On the other hand, for a more general case of a dissipative nanodevice presenting a potential $V_{ps}(z)$, we shall have different values of relaxation time for different energy levels.

For a monoenergetic particle beam, we have

$$\frac{p}{m} \frac{\partial}{\partial z} f_k(z,p) - \int dp' V_W(z, p-p') f_k(z,p') = -\frac{1}{\tau_k} \left( f_k(z,p) - f_{(k)eq}(z,p) \right), \quad (15)$$

where $f_{(k)eq}(z,p)$ is the equilibrium distribution resulting from the collisions between electrons with $k$ wave number and scattering centers.

It is important to note that, for practical cases, the energy dissipation varies in space so that its dependence in $z$ needs to be considered for a satisfactory description of the system. Such dependence shall results in non-local contributions to the collision term, which are very similar to what happens to the potential contribution (see Eqs. (10) and (11) ). However, Eqs. (14) and (15) are valid, for $\tau_k$ constant inside a certain region of space, if terms containing $\hbar/\tau_k$ are disconsidered. These terms are irrelevant for charge densities calculations, due to the space localization which arises from the integration over the entire momentum space.

The equation for $g_k(z,p) = f_k(z,p) - f_{(k)eq}(z,p)$ inside the device is given by

$$\frac{p}{m}\frac{\partial}{\partial z}g_k(z,p) - \int dp' V_W(z, p-p') g_k(z,p') = -\frac{1}{\tau_k}g_k(z,p). \qquad (16)$$

If $V(\mathbf{z})$ is a step potential, analytical solutions can be found by considering the break in the time symmetry due to energy dissipation. Such an effect can be mathematically described in the computation of Wigner function, Eq. (1), by replacing $k \to k + i\eta_k$ in the wave number, where

$$\eta_k = \frac{m}{2\tau_k \hbar k}, \qquad (17)$$

is a parameter which gives the intensity of energy dissipation.

The solutions generated by this way shall never exhibit unphysical feature such as the negative values in particle density early reported in literature for the conventional boundary conditions scheme [11]. General resulting profiles shall be that of attenuated beams of particles, with the addition of equilibrium distributions. A more profound analysis of this question shall be presented in a future work.

Therefore, the simplest situation in which this formalism can be applied is to the free particle case, where $V_W(z, p-p') = 0$ in Eq. (16). For the non-dissipative case, we have $\tau_k \to \infty$ so that $(p/m)\partial g_k / \partial z = 0$. The general solution with energy $\varepsilon_k = (\hbar k)^2 / 2m$ is given by

$$g_k(z,p) = C_1 \delta(p - \hbar k) + C_2 \delta(p + \hbar k) + \left[ C_3 \exp(-2izk)\delta(p) + c.c. \right]. \qquad (18)$$

where the integration constants $C_1$ and $C_2$ shall depends on the intensities of incident beans at the right and the left, respectively, and $C_3 = \sqrt{C_1 C_2} \exp(i\alpha)$, being $\alpha$ an arbitrary phase difference, shall be related to the entanglement between the incident and counter incident beams. The term "$c.c.$" represents the complex conjugate of the first expressions inside the brackets.

On the other hand, in the presence of dissipation, the solution for $g_k(z,p)$ shall split in two contributions, $g_{(k)c}(z,p)$, which is coherent, and $g_{(k)dc}(z,p)$ which is decoherent (i.e. the arbitrary phase difference ranging from $\alpha = 0$ to $\alpha = \pi$). The solution for $g_{(k)c}(z,p)$, being the dissipative region located between $z=a$ and $z=b$ (i.e. $l=b-a$), is given by

$$g_{c(k)}(z,p) = C_1 \exp(-2(z-a)\eta_k)\delta(p-\hbar k) + C_2 \exp(2(z-b)\eta_k)\delta(p+\hbar k) \\ + \left[ C_3 \exp(-2l\eta_k - 2izk)\tilde{\delta}(p+i\hbar\eta_k) + c.c. \right] \quad (19)$$

where the constants shall be determined by the electronic incidence at the boundaries, as in Eq. (18), and the function $\tilde{\delta}(p+i\hbar\eta_k)$ is the generalized delta function [10]. The solution for $g_{(k)dc}(z,p)$ shall be determined in combination to the equilibrium solution $f_{(k)eq}(z,p)$, in order to ensure the conservation of the number of particles which enters and leaves the dissipative region in a time unity.

The equilibrium function is a decoherent solution characterized by a temperature and a chemical potential. While temperature depends on some set of features such as the thermal bath and thermal capacities of the material, the chemical potential shall be the parameter effectively related to the number of particle in the system.

For sake of simplicity, let us consider the temperature as sufficiently low so that the Fermi-Dirac distribution falls in a Heaviside Theta Function, which means that all states bellow the Fermi level are occupied. In order to stablish the Fermi wave number $k_F$, we set the following assumptions: (i) the maximum value for $k_F$ is $k$, since collisions tends to lows energy (i.e. as $w \to \infty$, we shall have $k_F \to k$); (ii) the left leaving decoherent flux is equal to the right leaving decoherent flux. Consequently, we have

$$k_F = k\sqrt{1-\exp(-2l\eta_k)}, \quad (20)$$

where $l$ is the length of the dissipative region under consideration.

In this case, $f_{(k)eq}(z,p)$ can be written as

$$f_{(k)eq}(z,p) = \frac{C_1+C_2}{2}\int_0^{k_F} \frac{dq}{k}\left(\frac{\delta(p-\hbar q)+\delta(p+\hbar q)}{1-\exp(-2l\eta_q)}\right), \qquad (21)$$

with $k_F$ given by Eq. (20). The solution for $g_{(k)dc}(z,p)$ is stablished in order to ensure that thermalized particle beams are only leaving the region under consideration, and not entering it, which gives

$$g_{(k)dc}(z,p) = -\frac{C_1+C_2}{2}\int_0^{k_F} \frac{dq}{k}\left(\frac{\exp(-2(z-a)\eta_q)\delta(p-\hbar q)+\exp(2(z-b)\eta_q)\delta(p+\hbar q)}{1-\exp(-2l\eta_q)}\right). \qquad (22)$$

The general solution for Eq. (15), $f_k(z,p) = f_{(k)eq}(z,p) + g_{(k)dc}(z,p) + g_{(k)c}(z,p)$, yields

$$\begin{aligned}f_k(z,p) &= (C_1+C_2)\int_0^{k_F} \frac{dq}{k}\left[\frac{(1-\exp(-2(z-a)\eta_q))\delta(p-\hbar q)+(1-\exp(2(z-b)\eta_q))\delta(p+\hbar q)}{1-\exp(-2l\eta_q)}\right] \\ &+ C_1\exp(-2(z-a)\eta_k)\delta(p-\hbar k) + C_2\exp(2(z-b)\eta_k)\delta(p+\hbar k) \\ &+ \left[\sqrt{C_1 C_2}\exp(-2l\eta_k + i(\alpha-2zk))\tilde{\delta}(p+i\eta_k) + c.c.\right]\end{aligned} \qquad (23)$$

Eq. (23) has a quite clear physical meaning. The electron beam presenting a positive wave number becomes attenuated from the left to the right, meanwhile, the electron beam presenting a negative wave number becomes attenuated from the right to the left, which means that the deeper an electron penetrates in the dissipative region, the more it is likely to collide and fall in an arbitrary lower energy state.

It is also interesting to note that the entanglement term between positive and negative wave numbers (the last one in eq. (23)) is decreased by the existence of collisions. It shows that energy dissipation tends to obliterate quantum effects.

**Some Results and Comments**

In order to illustrate the applicability of the mothed developed above, lets us consider a monoenergetic beam of electrons with $50 meV$ injected in a dissipative nanodevice presenting a length of $40 nm$ with equal intensity by both sides. The density of particles along $z$ axis is given by the integration of $f_k(z,p)$, given by Eq. (23), over the entire $p$ space. The constants $C_1$ and $C_2$ are both set equal 1, in arbitrary unities, and with the phase difference $\alpha = 0$ between the coherent beams.

The results are shown in FIG. 1 for four different values of $\tau_k$. In the first one - FIG. 1(a) - we have $\tau_k = 5 ps$, which means a low dissipation of energy. This former profile results to be very similar to the "pure" quantum problem of a beam of free particles, resembling the known density profile obtained from the solution of Schrodinger Equation.

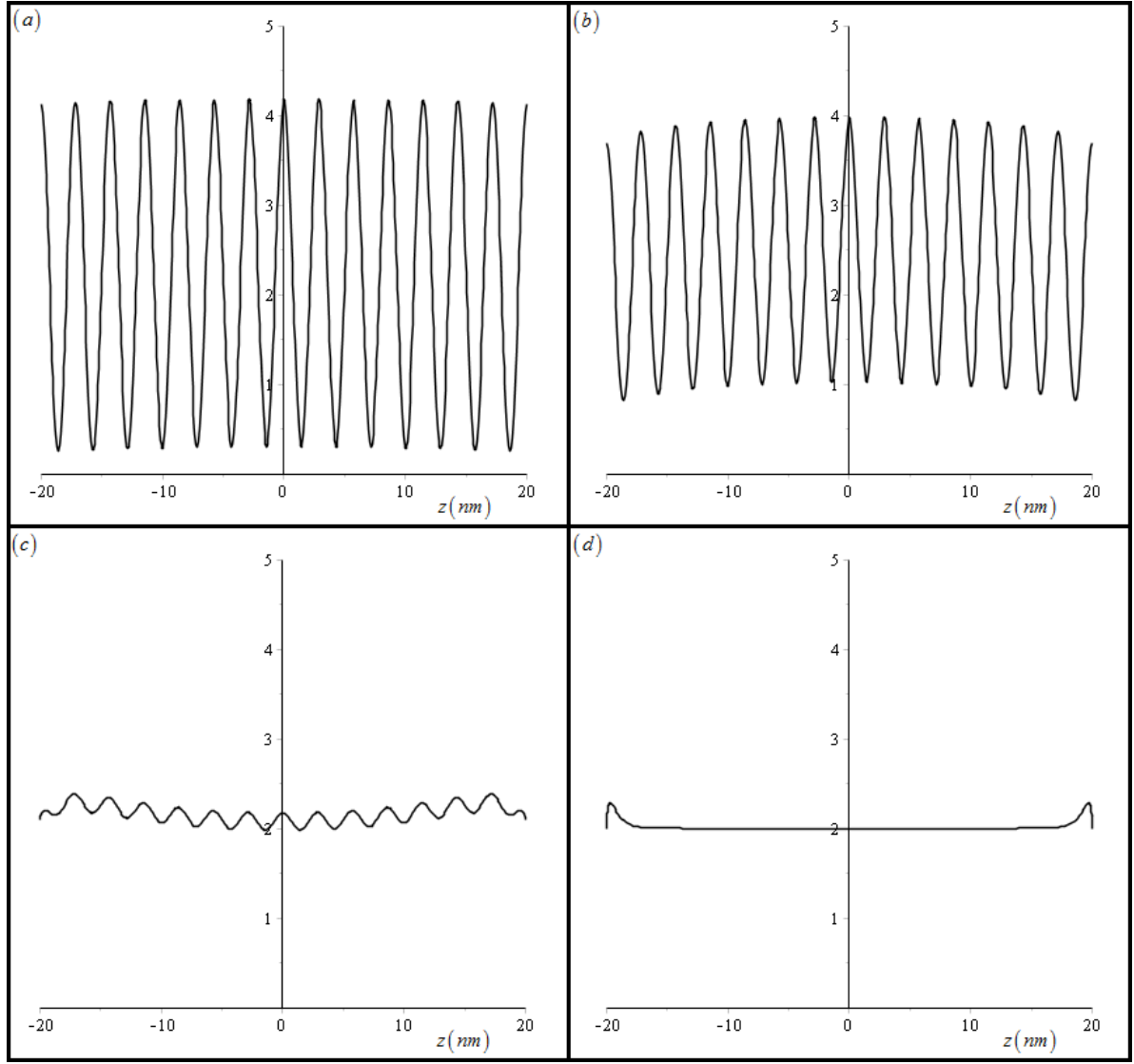

FIG1: Particle density profiles for a $50\,meV$ particle beam with relaxation time: (a) $\tau_k = 5\,ps$; (b) $\tau_k = 500\,fs$; (c) $\tau_k = 50\,fs$ and (d) $\tau_k = 5\,fs$ in arbitrary unities.

The other values are $\tau_k = 500\,fs$, $50\,fs$ and $5\,fs$, respectively. It is interesting to observe the vanishing of the quantum effects, which is characterized by the spatial oscillations in the particle density, as long as the relaxation time is decreased, i.e. the dissipation of energy is increased. Note that in FIG. 1(d) the decoherence dominates, and oscillations are practically eliminated.

These results can also be qualitatively interpreted in view of the mean penetration length $l_k$, which is defined as the mean length in which an electron penetrates the dissipative region before it suffer a collision. Its value shall be given by the product of the mean electron velocity and the relaxation time,

$$l_k = \frac{\hbar k}{m}\tau_k = \frac{1}{2\eta_k} . \qquad (24)$$

For $\tau_k = 5ps$, we have $l_k \simeq 663nm$, which is one order of magnitude higher than the length $l = 40nm$ of the nanodevice under consideration. It means that many of the electrons cross the region without collide, which explain why the density profile is very similar to that of non-dissipative case. Moreover, we observe that $l_k$ reduces one order of magnitude for each one of the other following values of $\tau_k$, so that for $\tau_k = 5fs$ we have $l_k \simeq 0,66nm$, being very unlikely for an electron to cross the region without collide.

**CONCLUSION**

We have developed an alternative approach to solve the Wigner-Boltzmann transport equation which is valid when relaxation time approximation is considered. An analytical solution was shown to exist for the case of step potential, which is of great importance in modelling nanodevises. An explicit solution was written for the case of free particle, in which a generalized delta function appears the entanglement term. The particle density profiles was also plotted for different values of relaxation time in order to illustrate the effect of energy dissipation on reduce quantum effects (decoherence). This approach is also capable to avoid certain unphysical features which arise from the conventional boundary condition scheme as early pointed out in the literature. A more rigorous analysis of this problem shall be presented in a future work.